\documentclass[aps,prd,nofootinbib,superscriptaddress,showpacs,floatfix,nofootinbib,amsmath,amssymb]
{revtex4}
\usepackage[dvips]{graphicx}

\begin{document}

\title{Exploring the Latest Union2 SNIa Dataset by Using Model-Independent Parametrization Methods}

\author{Shuang Wang}
\email{swang@mail.ustc.edu.cn} \affiliation{Department of Modern
Physics, University of Science and Technology of China, Hefei
230026, China} \affiliation{Institute of Theoretical Physics,
Chinese Academy of Sciences, Beijing 100190, China}

\author{Xiao-Dong Li}
\email{renzhe@mail.ustc.edu.cn} \affiliation{Interdisciplinary
Center for Theoretical Study, University of Science and Technology
of China, Hefei 230026, China} \affiliation{Institute of Theoretical
Physics, Chinese Academy of Sciences, Beijing 100190, China}

\author{Miao Li}
\email{mli@itp.ac.cn} \affiliation{Institute of Theoretical Physics,
Chinese Academy of Sciences, Beijing 100190, China}
\affiliation{Kavli Institute for Theoretical Physics China, Chinese
Academy of Sciences, Beijing 100190, China} \affiliation{Key
Laboratory of Frontiers in Theoretical Physics, Chinese Academy of
Sciences, Beijing 100190, China}

\begin{abstract}
We explore the cosmological consequences of the recently released Union2 sample of 557 Type Ia supernovae (SNIa).
Combining this latest SNIa dataset
with the Cosmic microwave background (CMB) anisotropy data from the Wilkinson Microwave Anisotropy Probe 7 year (WMAP7) observations
and the baryon acoustic oscillation (BAO) results from the Sloan Digital Sky Survey (SDSS) Data Release 7 (DR7),
we measure the dark energy density function $f(z)\equiv \rho_{de}(z)/\rho_{de}(0)$ as a free function of redshift.
Two model-independent parametrization methods
(the binned parametrization and the polynomial interpolation parametrization) are used in this paper.
By using the $\chi^2$ statistic and the Bayesian information criterion,
we find that the current observational data are still too limited to distinguish which parametrization method is better,
and a simple model has advantage in fitting observational data than a complicated model.
Moreover, it is found that all these parametrizations demonstrate that the Union2 dataset is still consistent with a cosmological constant at 1$\sigma$ confidence level.
Therefore, the Union2 dataset is different from the Constitution SNIa dataset, which more favors a dynamical dark energy.
\end{abstract}


\maketitle

\section{Introduction}\label{sec:intro}

Since the observations of type Ia supernovae (SNIa) first
indicated that the universe is undergoing  accelerated expansion at
the present stage \cite{Riess,Perlmutter}, dark energy (DE) has
become one of the most important problems in modern cosmology. Many
cosmologists believe that the cosmological constant fits the
observational data well. One also has reason to dislike the
cosmological constant since it poses the ¡°fine-tuning¡± and
¡°cosmic coincidence¡± puzzles \cite{Weinberg}. A variety of
proposals for dark energy have emerged, such as quintessence
\cite{quint}, phantom \cite{phantom}, $k$-essence \cite{k}, tachyon
\cite{tachyonic}, holographic \cite{holographic}, agegraphic
\cite{agegraphic}, hessence \cite{hessence}, Chaplygin gas
\cite{Chaplygin}, Yang-Mills condensate \cite{YMC}, etc.

A most powerful probe of DE is SNIa, which can be used as cosmological standard candles to measure directly the expansion history of the universe.
Recently, a large sample of SNIa, the Union2 SNIa dataset \cite{Union2}, was released.
This sample consists of 557 SNIa, covers a redshift region of $0\leq z\leq1.4$, and is the largest SNIa sample to date.
The Union2 dataset has been used to constrain various theoretical models \cite{WeiHao}.
However, to our best knowledge, this sample has not been analyzed by using a parametrization method that does not depend on any theoretical DE model.
What we shall do in this paper is just this.

Although constraining the equation of state parameter $w$ of DE is a popular and widely-used method to investigate DE,
Wang and Freese \cite{YunWang} pointed out that the DE density $\rho_{de}$ can be constrained more tightly than $w$ given the same observational data.
So in this paper,
combining the latest Union2 dataset
with the cosmic microwave background (CMB) anisotropy data from the Wilkinson Microwave Anisotropy Probe 7 years (WMAP7) observations \cite{WMAP7}
and the baryon acoustic oscillation (BAO) results from the Sloan Digital Sky Survey (SDSS) Data Release 7 (DR7) \cite{BAO},
we measure the DE density function $f(z)\equiv \rho_{de}(z)/\rho_{de}(0)$
as a free function of redshift.

Two model-independent parametrization methods are used in this
paper. First, we use the binned parametrization in which the
redshifts are separated into different bins and $f(z)$ is set as
constant in each bin. Binned fits of w have been applied before
\cite{Binned}, and similar analysis have been performed for density
binning \cite{Binned2}. In a previous work \cite{Huang}, we presented
a new binned parametrization method. Instead of setting the
discontinuity points of redshift by hand, we treated the
discontinuity points of redshift as free parameters and let them run
freely in the redshift range covered by SNIa data. As shown in
\cite{Huang}, this method can achieve much smaller $\chi_{min}^{2}$.
We shall use this new binned parametrization method to analyze the
Union2 sample.

Following \cite{YunWangParametrization},
we also consider the polynomial interpolation parametrization
in which $f(z)$ is parameterized by interpolating its own values at the redshifts $z_i=i*z_{max}/n ~ (i=1,2,...n)$.
Since the maximum redshift of the Union2 sample is 1.4,
in this paper we choose $z_{max}=1.4$, and consider the cases of $n=3$, $n=4$ and $n=5$.
As in \cite{YunWangParametrization}, we also set $f(z)$ as a constant in the range $z>1.4$
where DE is only weakly constrained by the CMB data.
Compared with the binned parametrization,
the advantage of the polynomial interpolation parametrization is that
the DE density function $f(z)$ can be reconstructed as a continuous function in the redshift range covered by SNIa data.

This paper is organized as follows.
In Sec. II, we describe the model-independent parametrizations considered here and the method of data analysis.
In Sec. III, we introduce the observational data and describe how they are included in our analysis.
In Sec. IV, we present the results obtained in this paper.
In the end, we give a short summary in Sec. V.
In this work, we assume today's scale factor $a_{0}=1$, so the redshift $z=a^{-1}-1$;
the subscript ``0'' always indicates the present value of the corresponding quantity, and the unit with $c=\hbar=1$ is used.

\section{Parametrizations And Methodology}

Standard candles impose constraints on cosmological parameters
through a comparison between the luminosity distance from observations and that from theoretical models.
In a spatially flat Friedmann-Robertson-Walker (FRW) universe (the assumption of flatness is motivated by the inflation scenario),
the luminosity distance $d_L$ is given by
\begin{equation}
\label{eq:dl}
d_L(z)=\frac{1+z}{H_{0}}\int_0^z\frac{dz'}{E(z')},
\end{equation}
with
\begin{equation}
\label{eq:Ez}
E(z)\equiv H(z)/H_{0} =\left[\Omega_{r}(1+z)^4+\Omega_m(1+z)^3+(1-\Omega_{r}-\Omega_{m})f(z)\right]^{1/2}.
\end{equation}
Here $H(z)$ is the Hubble parameter, $H_{0}$ is the Hubble constant, $\Omega_{m}$ is the present fractional matter density,
and $\Omega_{r}$ is the present fractional radiation density, given by \cite{WMAP7},
\begin{equation}
\Omega_{r}=\Omega_{\gamma}(1+0.2271N_{eff}),\ \ \
\Omega_{\gamma}=2.469\times10^{-5}h^{-2},\ \ \ N_{eff}=3.04,
\end{equation}
where $\Omega_{\gamma}$ is the present fractional photon density,
$h$ is the reduced Hubble parameter,
and $N_{eff}$ is the effective number of neutrino species.
Notice that the DE density function $f(z)\equiv \rho_{de}(z)/\rho_{de}(0)$ is a key function,
because a DE parametrization scheme enters in $f(z)$.

In the following we shall parameterize $f(z)$.
Notice that the maximum redshift of the Union2 SNIa sample is 1.4,
and in the range $z>1.4$ $f(z)$ is only weakly constrained by the CMB data.
As in \cite{YunWangParametrization}, we set $f(z)$ to be a constant in the range $z> 1.4$,
i.e.,
\begin{equation}
f(z)=f(1.4) ~~ (z>1.4).
\end{equation}
As pointed out in \cite{YunWangParametrization},
fixing $f(z>1.4)$ can help us to avoid making assumptions about early DE that can propagate
into artificial constraints on
DE at low $z$.

First, we use the binned parametrization, thus the DE density function $f(z)$ is parameterized as,
\begin{equation}
\label{eq:fzrbinned} f(z)= \left\{
\begin{array}{ll}
1 & 0\leq z \leq z_{1}
\\
\epsilon_{i} & z_{i-1} \leq z \leq z_{i} ~ (2\leq i\leq n)
\end{array}
\right..
\end{equation}
Here $\epsilon_{i}$ is a piecewise constant, and from the relation
$f(0)=1$ one can easily obtain $\epsilon_{1}=1$. It should be
mentioned that there are different opinions in the literature about
the optimal choice of redshift bins in constraining DE. In
\cite{Binned}, the authors directly set the discontinuity points
$z_i$ by hand. In \cite{Binned2}, Wang argues that one should choose
a constant $\Delta z$ for redshift slices. In this work, to have the
maximal freedom and to have the most model-independent
parametrizations, we do not fix the discontinuity points $z_i$ and
let them run freely in the region of $0\leq z\leq1.4$. As shown in
\cite{Huang}, this method can achieve much smaller $\chi_{min}^{2}$.
Here we consider the $n=2, 3$ cases.

Next, we use the method of polynomial interpolation to parameterize $f(z)$.
As in \cite{YunWangParametrization},
we choose  different redshift points $z_i=i*z_{max}/n ~ (i=1,2,...n)$,
and  interpolate $f(z)$ by using its own values at these redshift points.
This yields
\begin{equation}
f(z)=\sum_{i=1}^{n} f_i\frac{(z-z_1)...(z-z_{i-1})(z-z_{i+1})...(z-z_n)}{(z_i-z_1)...(z_i-z_{i-1})(z_i-z_{i+1})...(z_i-z_n)},
\end{equation}
where $f_i=f(z_i)$ and $z_n=z_{max}=1.4$.
Based on the relation $f(0)=1$, one parameter can be fixed directly (in this paper we fix $f_1$),
and only $n-1$ model parameters need to be determined by the data.
Here we consider the cases of $n=3$, $n=4$ and $n=5$.
Our parametrization is very similar to that of \cite{YunWangParametrization}.

In this work we adopt $\chi^2$ statistic \cite{swang} to estimate model parameters.
For a physical quantity $\xi$ with experimentally measured value $\xi_{obs}$,
standard deviation $\sigma_{\xi}$, and theoretically predicted value $\xi_{th}$,
$\chi^2$  is
\begin{equation}
\label{eq:chi2_xi}
\chi_{\xi}^2=\frac{\left(\xi_{obs}-\xi_{th}\right)^2}{\sigma_{\xi}^2}.
\end{equation}
The total $\chi^2$ is the sum of all $\chi_{\xi}^2$s,
i.e.
\begin{equation}
\label{eq:chi2}
\chi^2=\sum_{\xi}\chi_{\xi}^2.
\end{equation}
The best-fit model parameters are determined by minimizing the total
$\chi^{2}$. Moreover, by calculating $\Delta \chi^{2}\equiv
\chi^{2}-\chi_{min}^{2}$, one can determine the 1$\sigma$ and the
2$\sigma$ confidence level (CL) ranges of a specific model. Notice
that for the 1$\sigma$ and the 2$\sigma$ CL, different $n_p$ (denoting the number of
free model parameters) corresponds to different $\Delta
\chi^{2}$. Therefore, we list the
relationship between $n_p$ and $\Delta \chi^{2}$ in table \ref{01} from $n_p=1$ to $n_p=5$.

\begin{table}
\caption{\textrm{Relationship between number of free model
parameters $n_p$ and $\Delta \chi^{2}$ }}
\begin{center}
\label{01}
\begin{tabular}{|c|c|c|}
  \hline
  $n_p$ & $\Delta \chi^{2}$(1$\sigma$) & $\Delta \chi^{2}$(2$\sigma$) \\
  \hline
  1 & 1 & 4 \\
  \hline
  2 & 2.30 & 6.17 \\
  \hline
  3 & 3.53 & 8.02 \\
  \hline
  4 & 4.72 & 9.72 \\
  \hline
  5 & 5.89 & 11.3 \\
  \hline
\end{tabular}
\end{center}
\end{table}

For comparing different models, a statistical variable must be chosen.
The $\chi _{min}^{2}$ is the simplest one,
but it has difficulty to compare different models with different number of parameters.
In this work, we will use $\chi _{min}^{2}/dof$ as a model selection criterion,
where $dof$ is the degree of freedom defined as
\begin{equation}
\label{eq:dof}
dof\equiv N-n_{p},
\end{equation}
here $N$ is the number of data.
Besides, to compare different models with different number of parameters,
people often use the Bayesian information criterion \cite{Liddle} given by \cite{Schwarz}
\begin{equation}
\label{eq:BIC}
BIC=\chi_{min}^{2}+n_{p} \ln N.
\end{equation}
It is clear that a model favored by the observations should give smaller $\chi _{min}^{2}/dof$ and $BIC$.

\section{Observations}

First we start with the SNIa observations.
We use the latest Union2 sample including 557 data that are given in terms of the distance
modulus $\mu_{obs}(z_i)$ \cite{Union2}.
The theoretical distance modulus is defined as
\begin{equation}
\mu_{th}(z_i)\equiv 5 \log_{10} {D_L(z_i)} +\mu_0,
\end{equation}
where $\mu_0\equiv 42.38-5\log_{10}h$,
and in a flat universe the Hubble-free luminosity distance $D_L\equiv H_0 d_L$ ($d_L$ denotes the physical luminosity distance) is
\begin{equation}
D_L(z)=(1+z)\int_0^z {dz'\over E(z'; \theta)},
\end{equation}
where $\theta$ denotes the model parameters.
The $\chi^2$ for the SNIa data can be calculated as
\begin{equation}
\chi^2_{SN}(\theta)=\sum\limits_{i=1}^{557}{[\mu_{obs}(z_i)-\mu_{th}(z_i;\theta)]^2\over \sigma_i^2},\label{ochisn}
\end{equation}
where $\mu_{obs}(z_i)$ and $\sigma_i$ are the observed value and the corresponding 1$\sigma$ error of distance modulus for each supernova, respectively.
For convenient, people often analytically marginalize the nuisance parameter $\mu_0$
(i.e. the reduced Hubble constant $h$) when calculating $\chi^2_{SN}$ \cite{Perivolaropoulos}.

It should be stressed that the Eq.(\ref{ochisn}) only considers the statistical errors from SNIa,
and ignores the systematic errors from SNIa.
To include the effect of systematic errors into our analysis,
we will follow the prescription for using the Union2 compilation provided in \cite{Union Web}.
The key of this prescription is a $557 \times 557$ covariance matrix, $C_{SN}$,
which captures the systematic errors from SNIa (This covariance matrix with systematics can be downloaded from \cite{Union Web}).
Utilizing $C_{SN}$, we can calculate the following quantities
\begin{equation}
A=(\mu^{obs}_i-\mu^{th}_i)(C_{SN}^{-1})_{ij}(\mu^{obs}_j-\mu^{th}_j),\label{chisna}
\end{equation}
\begin{equation}
B=\sum\limits_{i=1}^{557}{(C_{SN}^{-1})_{ij}(\mu^{obs}_j-\mu^{th}_j)},\label{chisnb}
\end{equation}
\begin{equation}
C=\sum\limits_{i,j=1}^{557}{(C_{SN}^{-1})_{ij}},\label{chisnc}.
\end{equation}
Thus, the $\chi^2$ for the SNIa data is \cite{Union Web}
\begin{equation}
\chi^2_{SN}=A-\frac{B^{2}}{C}.\label{chisn}
\end{equation}
Different from the Eq.(\ref{ochisn}), this formula includes the effect of systematic errors from SNIa.

Then we turn to the CMB observations. Here we employ the ``WMAP
distance priors'' given by the 7-year WMAP observations
\cite{WMAP7}. This includes the ``acoustic scale'' $l_A$, the
``shift parameter'' $R$, and the redshift of the decoupling epoch of
photons $z_*$. The acoustic scale $l_A$ is defined as \cite{WMAP7}
\begin{equation}
\label{ladefeq} l_A\equiv (1+z_*){\pi D_A(z_*)\over r_s(z_*)}.
\end{equation}
Here $D_A(z)$ is the proper angular diameter distance, given by
\begin{equation}
D_A(z)=\frac{1}{1+z}\int^z_0\frac{dz^\prime}{E(z^\prime)},
\label{eq:da}
\end{equation}
and $r_s(z)$ is the comoving sound horizon size, given by
\begin{equation}
r_s(z)=\frac{1} {\sqrt{3}}  \int_0^{1/(1+z)}  \frac{ da } { a^2H(a)
\sqrt{1+(3\Omega_{b}/4\Omega_{\gamma})a} },
\label{eq:rs}
\end{equation}
where $\Omega_{b}$ and $\Omega_{\gamma}$ are the present baryon
and photon density parameters, respectively. In this paper, we adopt
the best-fit values, $\Omega_{b}=0.022765 h^{-2}$ and
$\Omega_{\gamma}=2.469\times10^{-5}h^{-2}$ (for $T_{cmb}=2.725$ K),
given by the 7-year WMAP observations \cite{WMAP7}. The fitting
function of $z_*$ is proposed by Hu and Sugiyama \cite{Hu:1995en}:
\begin{equation}
\label{zstareq} z_*=1048[1+0.00124(\Omega_b
h^2)^{-0.738}][1+g_1(\Omega_m h^2)^{g_2}],
\end{equation}
where
\begin{equation}
g_1=\frac{0.0783(\Omega_b h^2)^{-0.238}}{1+39.5(\Omega_b
h^2)^{0.763}},\quad g_2=\frac{0.560}{1+21.1(\Omega_b h^2)^{1.81}}.
\end{equation}
The shift parameter $R$ is defined as \cite{Bond97}
\begin{equation}
\label{shift} R(z_*)\equiv \sqrt{\Omega_m H_0^2}(1+z_*)D_A(z_*).
\end{equation}
Following Ref.\cite{WMAP7}, we use the prescription for using the
WMAP distance priors. Thus, the $\chi^2$ for the CMB data is
\begin{equation}
\chi_{CMB}^2=(x^{obs}_i-x^{th}_i)(C_{CMB}^{-1})_{ij}(x^{obs}_j-x^{th}_j),\label{chicmb}
\end{equation}
where $x_i=(l_A, R, z_*)$ is a vector, and $(C_{CMB}^{-1})_{ij}$ is the
inverse covariance matrix. The 7-year WMAP observations \cite{WMAP7}
give the maximum likelihood values: $l_A(z_*)=302.09$,
$R(z_*)=1.725$, and $z_*=1091.3$. The inverse covariance matrix is
also given in Ref. \cite{WMAP7}
\begin{equation}
(C_{CMB}^{-1})=\left(
  \begin{array}{ccc}
    2.305 & 29.698 & -1.333 \\
    29.698 & 6825.27 & -113.180 \\
    -1.333 & -113.180  &  3.414 \\
  \end{array}
\right).
\end{equation}

At last we consider the BAO observations.
The spherical average gives us the following effective distance measure \cite{Eisenstein}
\begin{equation}
 D_V(z) \equiv \left[(1+z)^2D_A^2(z)\frac{z}{H(z)}\right]^{1/3},
\end{equation}
where $D_A(z)$ is the proper angular diameter distance given in Eq.(\ref{eq:da}).
As in \cite{BAO}, we focus on a quantity $r_s(z_d)/D_V(0.275)$,
where $r_s$ is given in Eq.(\ref{eq:rs}),
and $z_d$ denotes the redshift of the drag epoch, whose fitting formula is proposed by Eisenstein and Hu \cite{BAODefzd}
\begin{equation}
\label{Defzd} z_d={1291(\Omega_mh^2)^{0.251}\over 1+0.659(\Omega_mh^2)^{0.828}}\left[1+b_1(\Omega_bh^2)^{b2}\right].
\end{equation}
where
\begin{equation}
\label{Defb1b2} b_1=0.313(\Omega_mh^2)^{-0.419}\left[1+0.607(\Omega_mh^2)^{0.674}\right], \quad  b_2=0.238(\Omega_mh^2)^{0.223}.
\end{equation}
The BAO data from the spectroscopic SDSS DR7 galaxy sample galaxy sample \cite{BAO} give $r_s(z_d)/D_V(0.275)=0.1390\pm 0.0037$.
Thus, the $\chi^2$ for the BAO data is,
\begin{equation}
\chi^2_{BAO}=\left(\frac{r_s(z_d)/D_V(0.275)-0.1390}{0.0037}\right)^2.
\end{equation}

\section{Results}

\subsection{Binned Parametrization}

In this subsection we will discuss the binned parametrization. As
mentioned above, we will consider the cases of $n=2$ and $n=3$. For
simplicity, we will call 2 bins piecewise constant $\rho_{de}$
parametrization the $\Lambda$CDM2 model, and will call 3 bins
piecewise constant $\rho_{de}$ parametrization  the$\Lambda$CDM3
model.

\begin{figure}
\includegraphics[scale=0.7, angle=0]{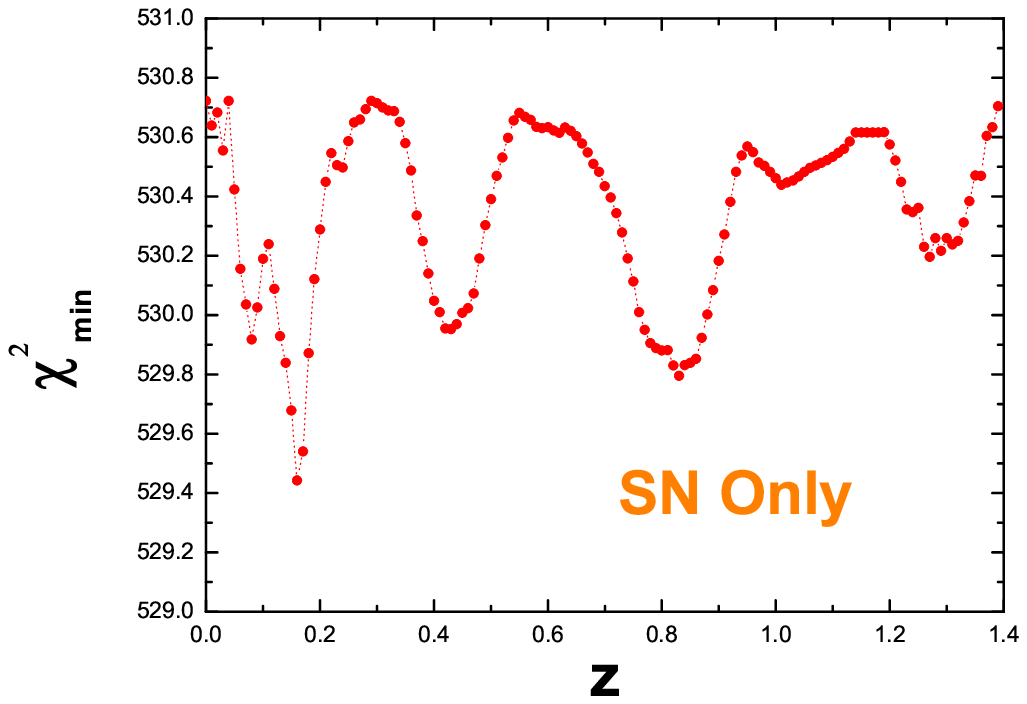}
\includegraphics[scale=0.7, angle=0]{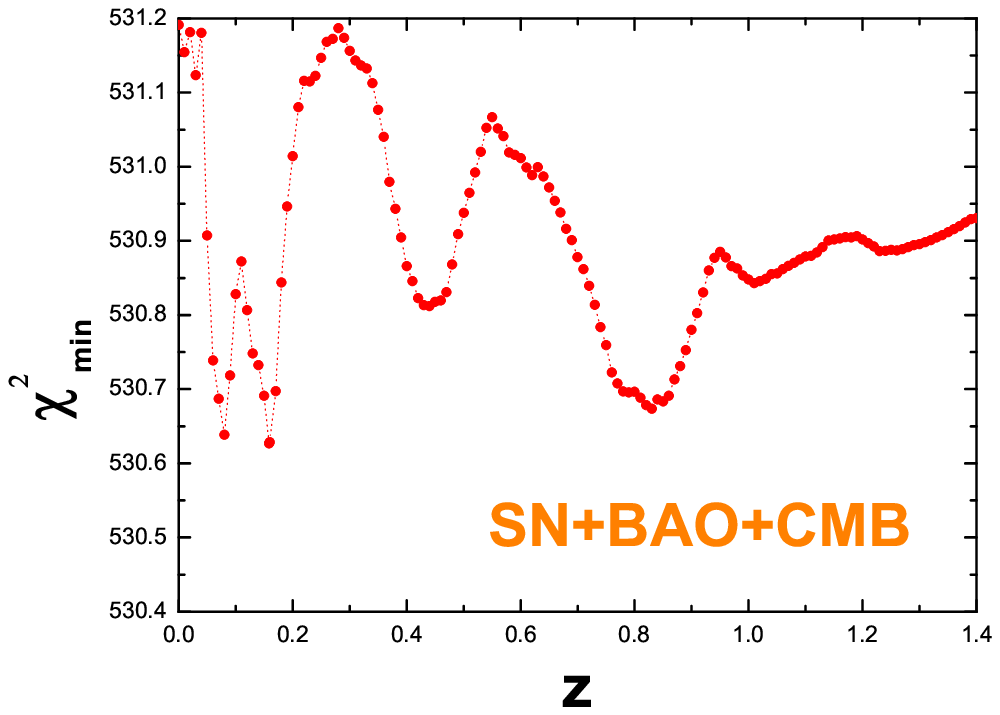}
\caption{$\chi_{min}^{2}$ versus redshift $z$ for the $\Lambda$CDM2 models.
The left panel is plotted by using the Union2 SNIa sample alone,
and the right panel is plotted by using the combined SNIa+CMB+BAO data.} \label{Fig:01}
\end{figure}

First we discuss the $\Lambda$CDM2 model.
Figure \ref{Fig:01} shows $\chi_{min}^{2}$ versus redshift $z$ for the $\Lambda$CDM2 model,
where the Union2 sample and the combined SNIa+CMB+BAO data are used, respectively.
It is found that using Union2 alone, the $\Lambda$CDM2 model achieves its minimal $\chi_{min}^{2}=529.43$ when $z_1 = 0.162$,
while the best-fit value and the corresponding 1$\sigma$ CL of the model parameters are $\Omega_{m}=0.326_{-0.089}^{+0.101}$ and $\epsilon_{2}=0.847_{-0.250}^{+0.201}$.
Using the combined SNIa+CMB+BAO data, the $\Lambda$CDM2 model achieves its minimal
$\chi_{min}^{2}=530.627$ when $z_1 = 0.158$,
while the best-fit value and the corresponding 1$\sigma$ CL of the model parameters are $\Omega_{m}=0.279_{-0.024}^{+0.026}$ and $\epsilon_{2}=0.947_{-0.100}^{+0.110}$.

\begin{figure}
\includegraphics[scale=0.7, angle=0]{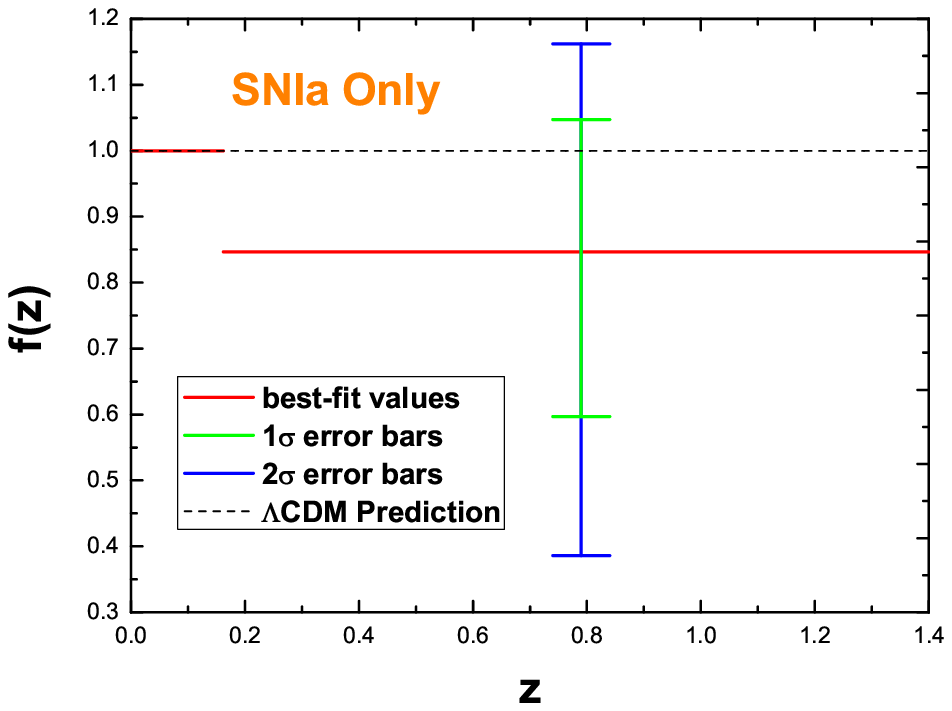}
\includegraphics[scale=0.7, angle=0]{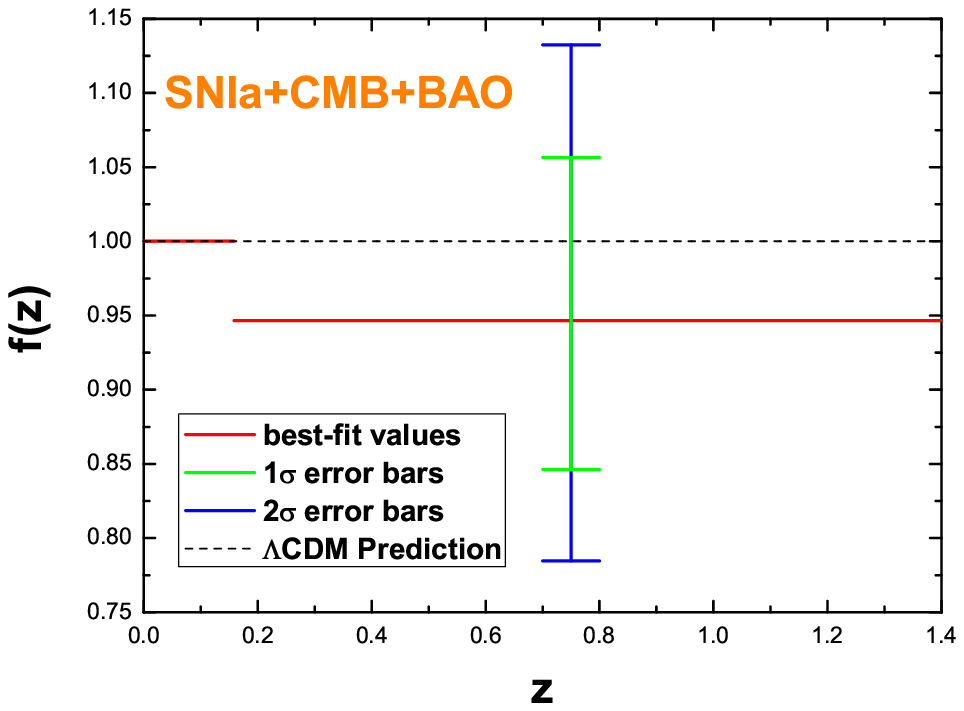}
\caption{The evolution of $f(z)$ along with $z$ for the $\Lambda$CDM2 model.
The left panel is plotted by using the Union2 SNIa sample alone,
and the right panel is plotted by using the combined SNIa+CMB+BAO data.} \label{Fig:02}
\end{figure}

In figure \ref{Fig:02}, we plot the evolution of $f(z)$ along with
$z$ for the $\Lambda$CDM2 model.
Based on the best-fit results shown in this figure,
it is found that the Union2 dataset favors a decreasing $f(z)$,
while the combined SNIa+CMB+BAO data favor a slowly decreasing $f(z)$.
This means that, compared with the result given by the SNIa data alone,
the result of the combined SNIa+CMB+BAO data is more close to the $\Lambda$CDM model (i.e. the cosmological constant model).
Moreover, after taking into account the error bars,
the Union2 dataset is consistent with the $\Lambda$CDM model at 1$\sigma$ CL.

\begin{figure}
\includegraphics[scale=0.4, angle=0]{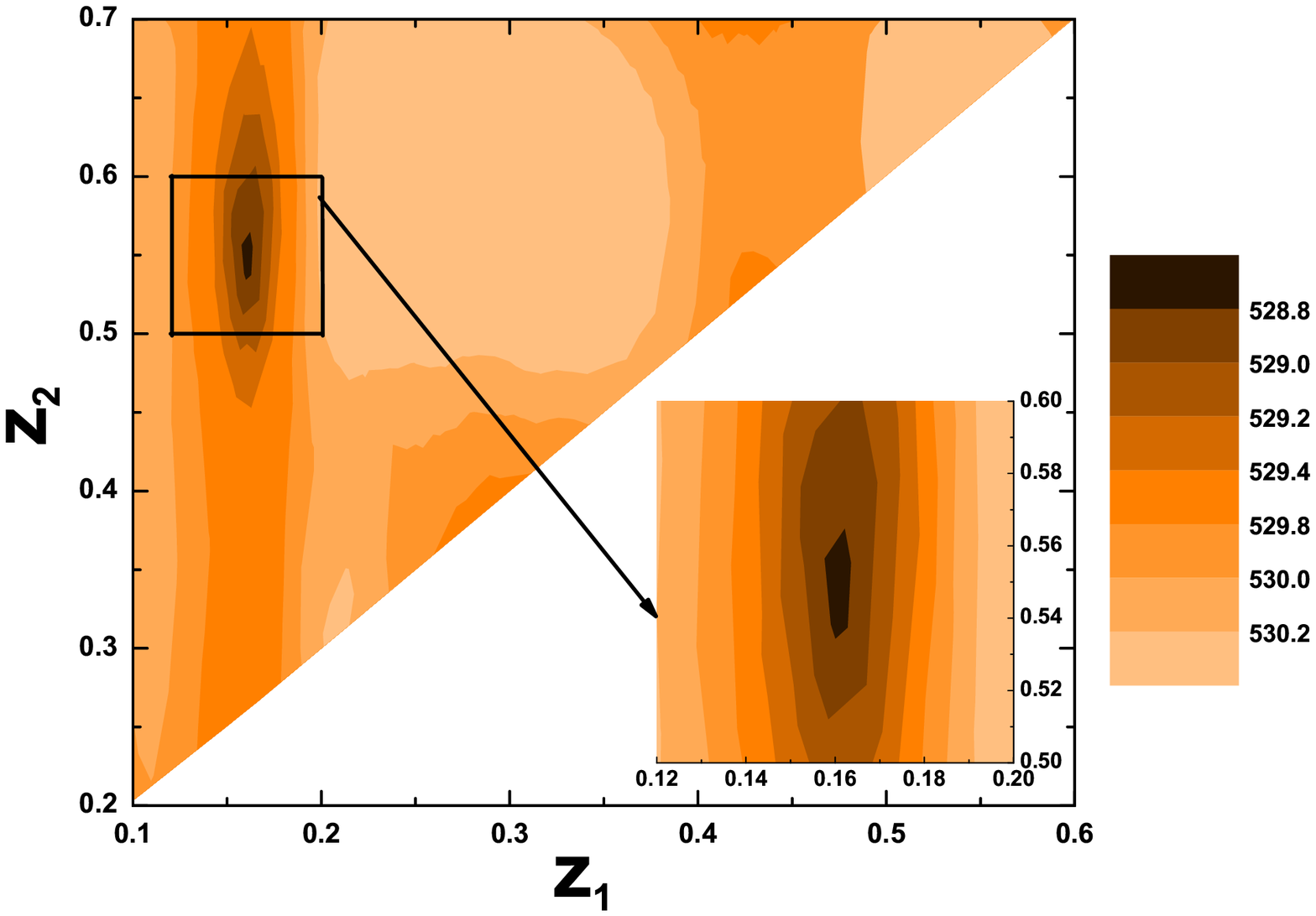}
\includegraphics[scale=0.4, angle=0]{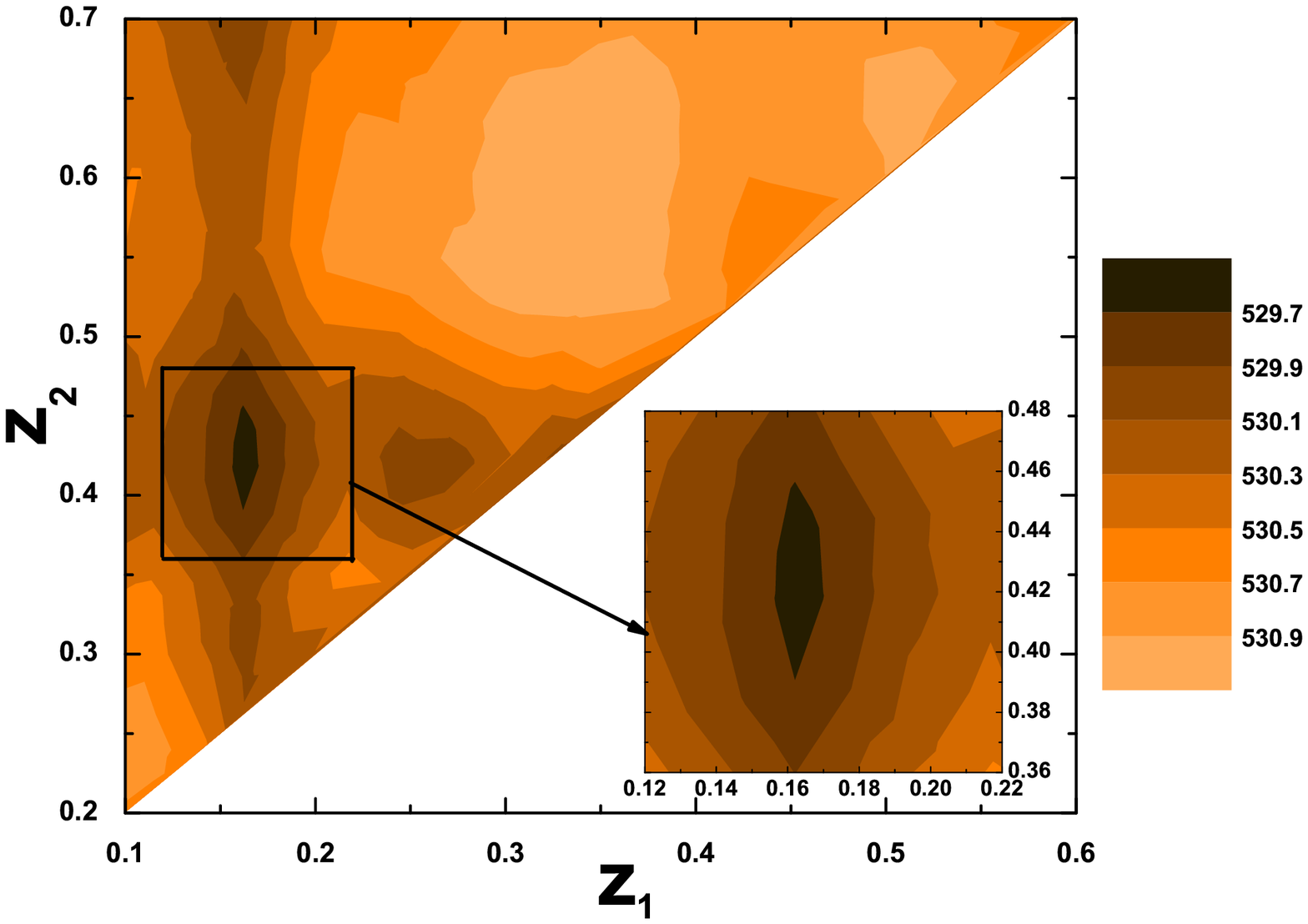}
\caption{The relationship between the $\chi_{min}^{2}$
and the discontinuity points of redshift ($z_1$ and $z_2$) for the $\Lambda$CDM3 model.
The left panel is plotted by using the Union2 SNIa sample alone,
and the right panel is plotted by using the combined SNIa+CMB+BAO data.
The x-axis represents the redshift of the first discontinuity point $z_1$,
while the y-axis denotes the redshift of the second discontinuity point $z_2$.
Notice that the light-colored region corresponds to a big $\chi^{2}$,
and the dark-colored region corresponds to a small $\chi^{2}$.
Since $z_1\leq z_2$ must be satisfied,
the bottom-right region of the figure is always blank.} \label{Fig:03}
\end{figure}

Next we turn to the $\Lambda$CDM3 model.
Figure \ref{Fig:03} shows the relationship between the $\chi_{min}^{2}$ and the discontinuity points of redshift ($z_1$ and $z_2$) for the $\Lambda$CDM3 models,
where the Union2 sample and the combined SNIa+CMB+BAO data are used, respectively.
It is found that using the Union2 dataset alone, the $\Lambda$CDM3 model achieves its minimal $\chi_{min}^{2}=528.621$
when $z_1 = 0.162$ and $z_2 = 0.552$,
while the best-fit value and the corresponding 1$\sigma$ CL of the model parameters are
$\Omega_{m}=0.427_{-0.231}^{+0.266}$, $\epsilon_{2}=0.665_{-1.083}^{+0.407}$ and $\epsilon_{3}=-0.130_{-4.506}^{+1.766}$.
Using the combined SNIa+CMB+BAO data, the $\Lambda$CDM3 model achieves its minimal $\chi_{min}^{2}=529.622$
when $z_1 = 0.162$ and $z_2 = 0.421$,
while the best-fit value and the corresponding 1$\sigma$ CL of the model parameters are
$\Omega_{m}=0.276_{-0.029}^{+0.033}$, $\epsilon_{2}=0.915_{-0.131}^{+0.150}$ and $\epsilon_{3}=1.091_{-0.294}^{+0.335}$.

\begin{figure}
\includegraphics[scale=0.7, angle=0]{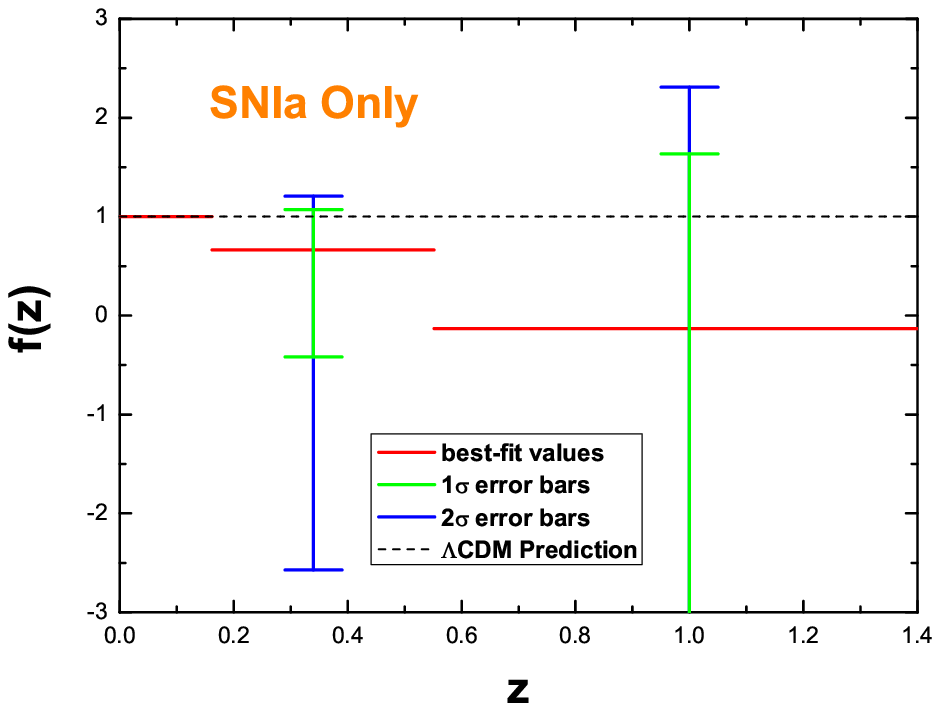}
\includegraphics[scale=0.7, angle=0]{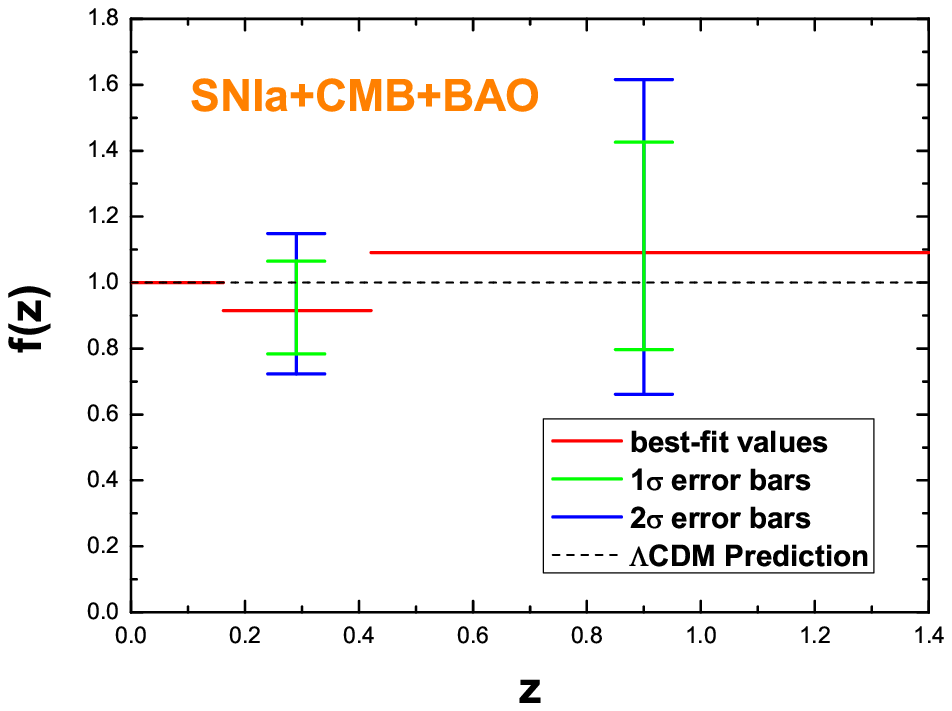}
\caption{The evolution of $f(z)$ along with $z$ for the $\Lambda$CDM3 model.
The left panel is plotted by using the Union2 SNIa sample alone,
and the right panel is plotted by using the combined SNIa+CMB+BAO data.} \label{Fig:04}
\end{figure}

In figure \ref{Fig:04}, we plot the evolution of $f(z)$ along with $z$ for the $\Lambda$CDM3 model.
Based on the best-fit results shown in this figure,
it is found that the Union2 dataset favors a decreasing $f(z)$,
while the combined SNIa+CMB+BAO data favor an oscillating $f(z)$, which is more close to the $\Lambda$CDM model.
Moreover, after taking into account the error bars,
one can see that the Union2 dataset is consistent with the $\Lambda$CDM model at 1$\sigma$ CL.

\begin{figure}
\includegraphics[scale=0.7, angle=0]{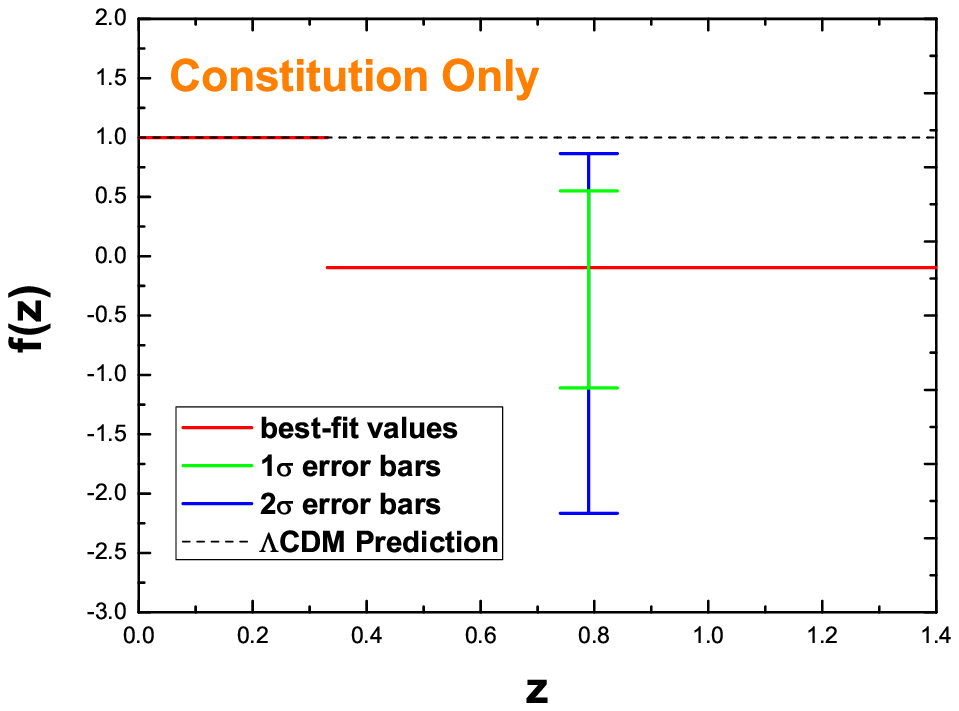}
\includegraphics[scale=0.7, angle=0]{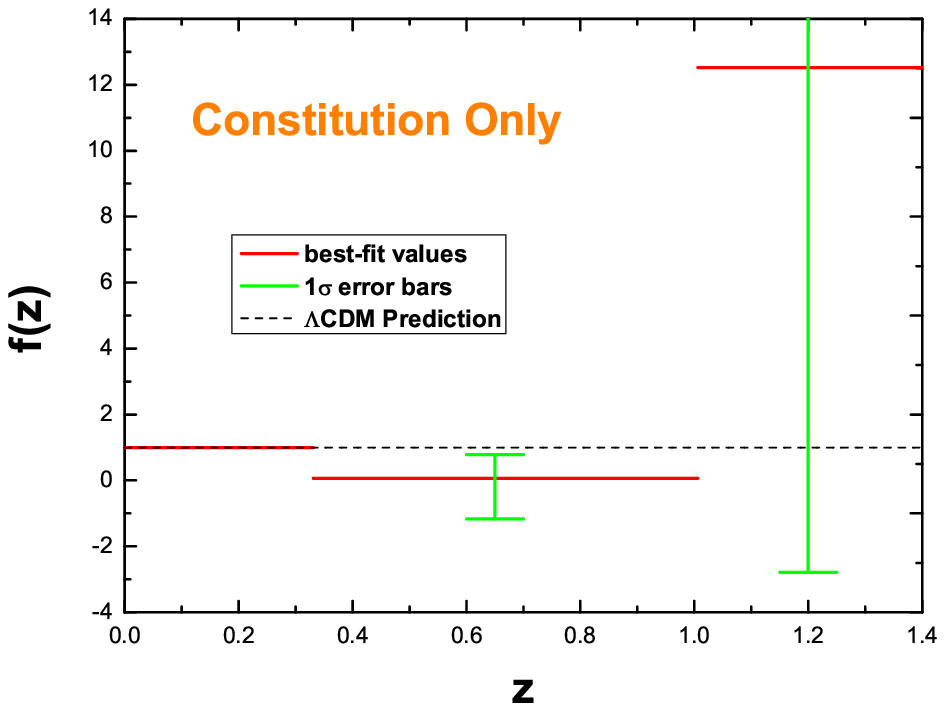}
\caption{The evolution of $f(z)$ along with $z$ given by the Constitution dataset alone.
The left panel is the case of the $\Lambda$CDM2 model,
and the right panel is the case of the $\Lambda$CDM3 model.} \label{Fig:05}
\end{figure}

As a comparison, for the $\Lambda$CDM2 the $\Lambda$CDM3 model,
we also plot the evolution of $f(z)$ given by the Constitution dataset \cite{Hicken2} alone in figure \ref{Fig:05}.
As seen in this figure, for the $\Lambda$CDM2 model,
there is a deviation from the $\Lambda$CDM model at 2$\sigma$ CL,
and for the $\Lambda$CDM3 model,
there is a deviation from the $\Lambda$CDM model at 1$\sigma$ CL.
This means that the Constitution sample more favors a dynamical DE.
Therefore, the Union2 dataset is evidently different from the Constitution dataset.

\subsection{Polynomial Interpolation Parametrization}

In this subsection we will discuss the polynomial interpolation parametrization.
Compared with the binned parametrization,
the advantage of the polynomial interpolation parametrization is that
the DE density function $f(z)$ can be reconstructed as a continuous function in the redshift range covered by SNIa data.
As mentioned above, we will consider the  $n=3, 4, 5$ cases.
For simplicity, we will call the polynomial interpolation parametrization of $n=3$ the PI3 model,
will call the polynomial interpolation parametrization of $n=4$ the PI4 model,
and will call the polynomial interpolation parametrization of $n=5$ the PI5 model.

\begin{figure}
\includegraphics[scale=0.7, angle=0]{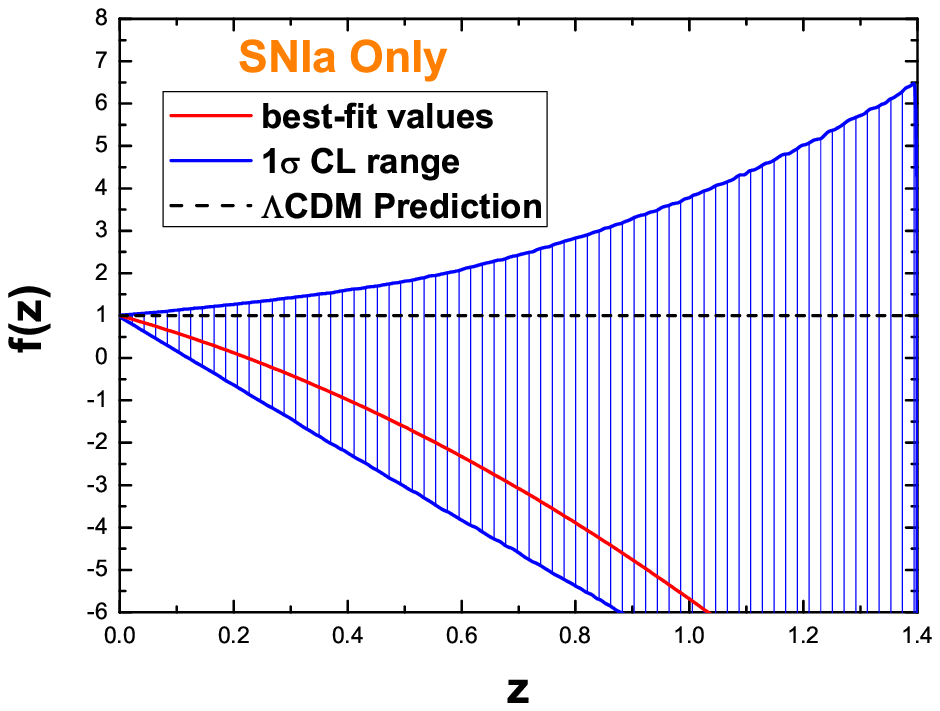}
\includegraphics[scale=0.7, angle=0]{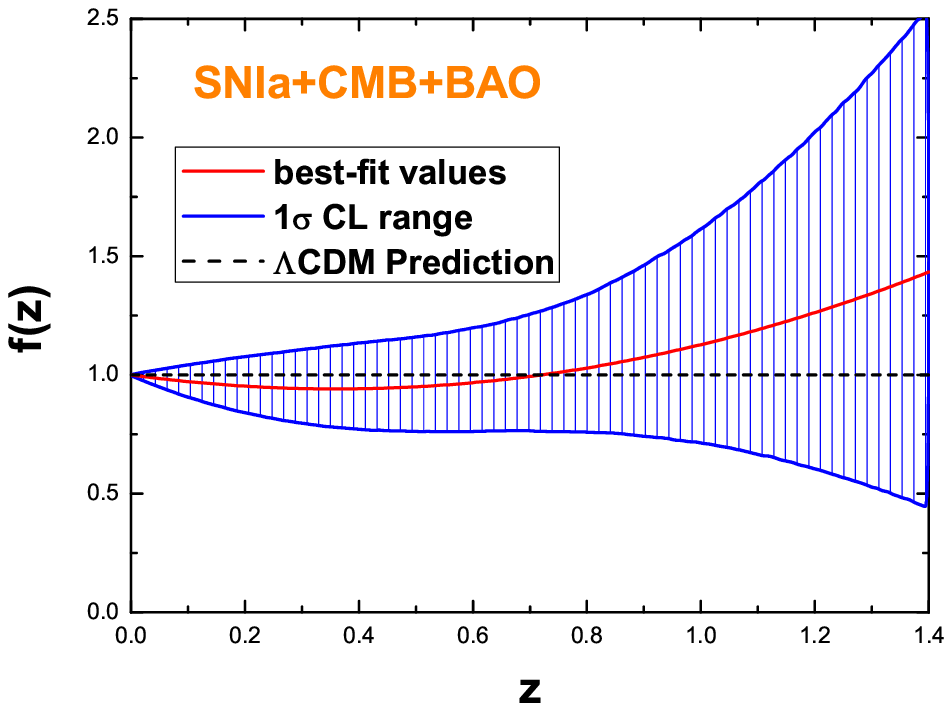}
\caption{The evolution of $f(z)$ along with $z$ for the PI3 model.
The left panel is plotted by using the Union2 SNIa sample alone,
and the right panel is plotted by using the combined SNIa+CMB+BAO data.} \label{Fig:06}
\end{figure}

First we discuss the PI3 model.
Using the Union2 dataset alone, the PI3 model has a minimal $\chi_{min}^{2}=530.102$,
while the best-fit value of the model parameters are $\Omega_{m}=0.650$, $f_{2}=-5.063$ and $f_{3}=-9.970$.
Using the combined SNIa+CMB+BAO data, the PI3 model has a minimal $\chi_{min}^{2}=530.391$,
while the best-fit value of the model parameters are $\Omega_{m}=0.279$, $f_{2}=1.091$ and $f_{3}=1.434$.
In figure \ref{Fig:06}, we plot the evolution of $f(z)$ along with $z$ for the PI3 model.
Based on the best-fit results shown in this figure,
it is found that the Union2 dataset favors a rapidly decreasing $f(z)$,
while the combined SNIa+CMB+BAO data favor an oscillating $f(z)$, which is more close to the $\Lambda$CDM model.
Moreover, after taking into account the error bars,
the Union2 dataset is still consistent with the $\Lambda$CDM model at 1$\sigma$ CL.
These results are similar to the results of figure \ref{Fig:04}.

\begin{figure}
\includegraphics[scale=0.7, angle=0]{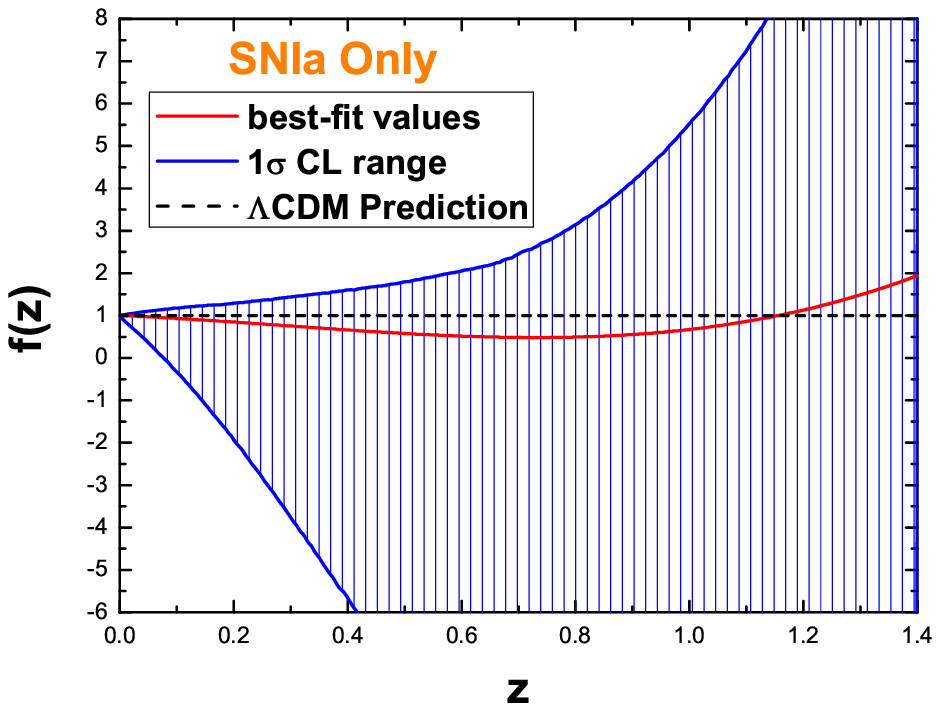}
\includegraphics[scale=0.7, angle=0]{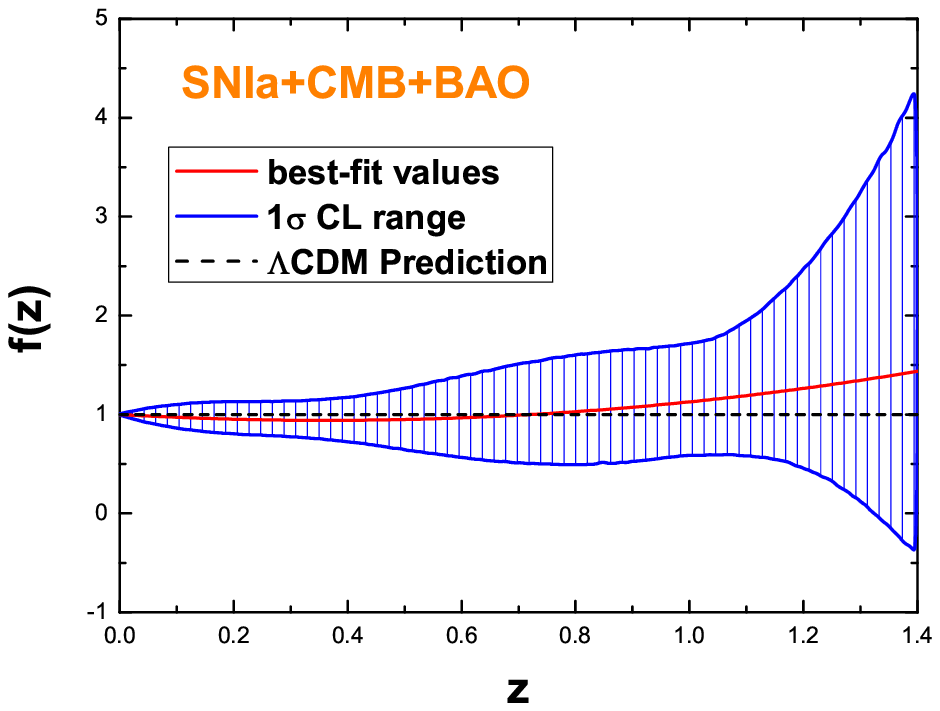}
\caption{The evolution of $f(z)$ along with $z$ for the PI4 model.
The left panel is plotted by using the Union2 SNIa sample alone,
and the right panel is plotted by using the combined SNIa+CMB+BAO data.} \label{Fig:07}
\end{figure}

Next we turn to the PI4 model.
Using the Union2 dataset alone, the PI4 model has a minimal $\chi_{min}^{2}=530.031$,
while the best-fit value of the model parameters are $\Omega_{m}=0.368$, $f_{2}=0.485$, $f_{3}=0.758$ and $f_{4}=1.944$.
Using the combined SNIa+CMB+BAO data, the PI4 model has a minimal $\chi_{min}^{2}=530.390$,
while the best-fit value of the model parameters are $\Omega_{m}=0.279$, $f_{2}=0.993$, $f_{3}=1.158$ and $f_{4}=1.437$.
In figure \ref{Fig:07}, we plot the evolution of $f(z)$ along with $z$ for the PI4 model.
Based on the best-fit results shown in this figure,
it is found that both the Union2 dataset alone and the combined SNIa+CMB+BAO data favor an oscillating $f(z)$,
while the result of the combined SNIa+CMB+BAO data is more close to the $\Lambda$CDM model.
Moreover, after taking into account the error bars,
the Union2 dataset is still consistent with the $\Lambda$CDM model at 1$\sigma$ CL.

\begin{figure}
\includegraphics[scale=0.7, angle=0]{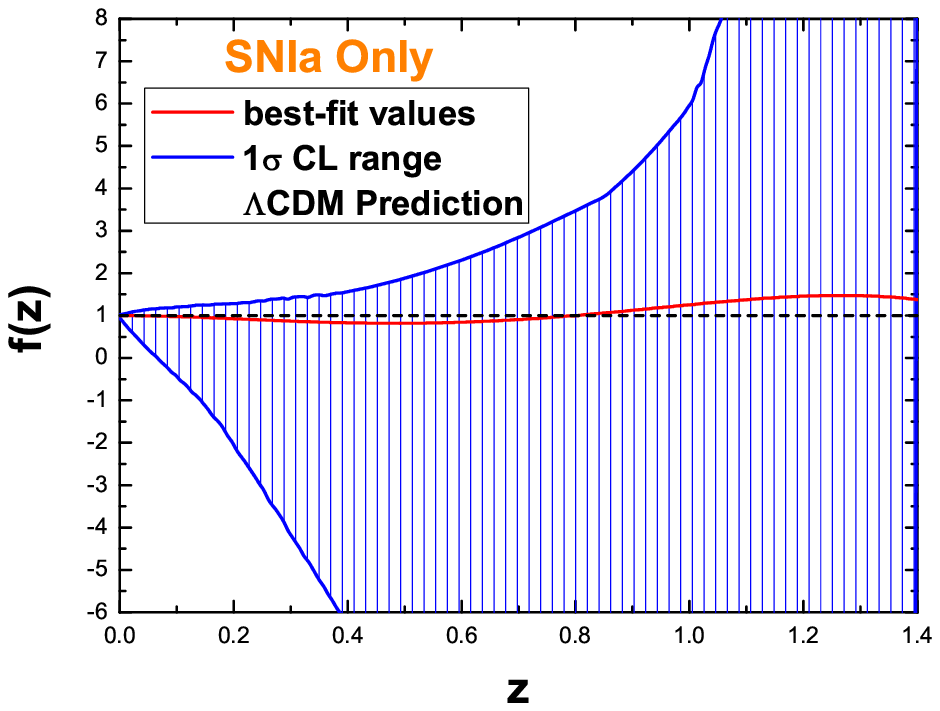}
\includegraphics[scale=0.7, angle=0]{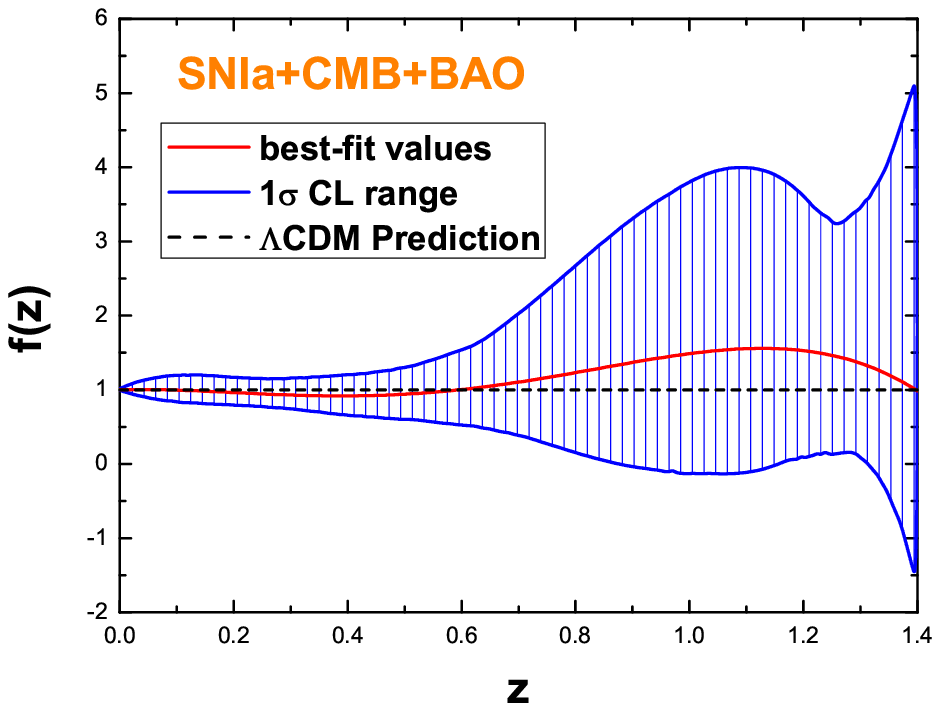}
\caption{The evolution of $f(z)$ along with $z$ for the PI5 model.
The left panel is plotted by using the Union2 SNIa sample alone,
and the right panel is plotted by using the combined SNIa+CMB+BAO data.} \label{Fig:08}
\end{figure}

Lately we consider the PI5 model.
Using the Union2 dataset alone, the PI5 model has a minimal $\chi_{min}^{2}=529.994$,
while the best-fit values of the model parameters are $\Omega_{m}=0.312$, $f_{2}=0.829$, $f_{3}=1.048$, $f_{4}=1.394$ and $f_{5}=1.375$.
Using the combined SNIa+CMB+BAO data, the PI5 model has a minimal $\chi_{min}^{2}=530.145$,
while the best-fit values of the model parameters are $\Omega_{m}=0.280$, $f_{2}=0.973$, $f_{3}=1.286$, $f_{4}=1.558$ and $f_{5}=0.984$.
In figure \ref{Fig:08}, we plot the evolution of $f(z)$ along with $z$ for the PI5 model.
This figure has some subtle differences with the figure \ref{Fig:07}, but
both the Union2 dataset alone and the combined SNIa+CMB+BAO data still favor an oscillating $f(z)$.
Again, after taking into account the error bars, one can see that the Union2 dataset is consistent with the $\Lambda$CDM model at 1$\sigma$ CL.
Therefore, based on figure \ref{Fig:06}, figure \ref{Fig:07} and figure \ref{Fig:08},
one can see that the polynomial interpolation parametrization also demonstrate that
the Union2 dataset is still consistent with a cosmological constant at 1$\sigma$ CL.

\subsection{A Comparison of Parametrization Methods}

In this subsection we will make a comparison of the parametrization methods considered in this work.
Utilizing the combined SNIa+CMB+BAO data,
we list the $\chi_{min}^{2}$, the $\chi_{min}^{2}/dof$ and the $BIC$ for these models in table \ref{02}.
Based on this table, two conclusions can be obtain.
First, the differences among these models' $\chi_{min}^{2}$ are very small.
Therefore, the current observational data are still too limited to distinguish which parametrization method is better.
Second, the $\Lambda$CDM2 model has lower $\chi_{min}^{2}/dof$ and $BIC$ than the $\Lambda$CDM3 model,
and the cases for the polynomial interpolation parametrization are similar.
This hints that a simple model has advantage in fitting observational data than a complicated model.
Moreover, we also include the $\Lambda$CDM model in this table for comparison.
It is found that a cosmological constant is still most favored by the cosmological observations.
This result is consistent with the conclusions of our paper.

\begin{table}
\caption{\textrm{A comparison of parametrization methods, where the combined SNIa+CMB+BAO data are used during the analysis.
$n_p$ is the number of free model parameters. We also include the $\Lambda$CDM model in this table for comparison.}}
\begin{center}
\label{02}
\begin{tabular}{|c|c|c|c|c|}
  \hline
  ~ Model ~ & ~~ $n_p$ ~~ & ~ $\chi_{min}^{2}$ ~ & $\chi_{min}^{2}/dof$ & ~ $BIC$ ~ \\
  \hline
  $\Lambda$CDM & 1 & 531.192 & 0.949 & 537.522 \\
  \hline
  $\Lambda$CDM2 & 3 & 530.627 & 0.951 & 549.616 \\
  \hline
  $\Lambda$CDM3 & 5 & 529.622 & 0.953 & 561.271 \\
  \hline
  PI3 & 3 & 530.391 & 0.951 & 549.380 \\
  \hline
  PI4 & 4 & 530.390 & 0.952 & 555.710 \\
  \hline
  PI5 & 5 & 530.145 & 0.953 & 561.794 \\
  \hline
\end{tabular}
\end{center}
\end{table}

\section{Summary}

In this work, we explore the constraint of the recently released Union2 sample of 557 SNIa on DE.
Combining this latest SNIa dataset with the BAO results from the SDSS measurements and the CMB anisotropy data from the WMAP7 observations,
we measure the DE density function $f(z)\equiv \rho_{de}(z)/\rho_{de}(0)$ as a free function of redshift.
To extract information directly from current observational data,
two model-independent parametrization methods are used in this paper.
By using the $\chi^2$ statistic and the Bayesian information criterion,
we find that the current observational data are still too limited to distinguish which parametrization method is better,
and a simple model has advantage in fitting observational data than a complicated model.
Moreover, it is found that all these parametrizations demonstrate that the Union2 dataset is still consistent with a cosmological constant at 1$\sigma$ confidence level.
Therefore, the Union2 dataset is different from the Constitution SNIa dataset, which more favors a dynamical dark energy.

\begin{acknowledgments}
We are grateful to the referee for helpful suggestions.
This work was supported by the NSFC grant No.10535060/A050207, a NSFC group grant No.10821504
and Ministry of Science and Technology 973 program under grant No.2007CB815401.
Shuang Wang was also supported by a graduate fund of USTC.
\end{acknowledgments}



\begin{thebibliography}{99}


\bibitem{Riess}
A.G. Riess  et al., AJ. {\bf 116}, 1009  (1998).

\bibitem{Perlmutter}
S. Perlmutter et al., ApJ {\bf 517}, 565 (1999).


\bibitem{Weinberg} S. Weinberg, Rev. Mod. Phys. {\bf 61}, 1 (1989);
V. Sahni and A.A. Starobinsky, Int. J. Mod. Phys. D {\bf 9}, 373 (2000);
S.M. Carroll, Living Rev.Rel. {\bf 4}, 1 (2001);
P.J.E. Peebles and B. Ratra, Rev. Mod. Phys. {\bf 75}, 559 (2003);
T. Padmanabhan, Phys. Rept. {\bf 380}, 235 (2003);
E.J. Copeland, M. Sami and S. Tsujikawa, Int. J. Mod. Phys. D {\bf 15}, 1753 (2006).


\bibitem{quint} B. Ratra and P.J.E. Peebles, Phys. Rev. D{\bf 37}, 3406 (1988);
P.J.E. Peebles and B.Ratra,  ApJ {\bf325}, L17 (1988);
C. Wetterich, Nucl. Phys. B{\bf 302}, 668 (1988);
I. Zlatev, L. Wang and P.J. Steinhardt, Phys. Rev. Lett. {\bf 82}, 896 (1999).


\bibitem{phantom}
R.R. Caldwell, Phys. Lett. B {\bf 545}, 23 (2002); S.M. Carroll, M.
Hoffman and M. Trodden, Phys. Rev. D{\bf 68}, 023509 (2003).

\bibitem{k}
C. Armendariz-Picon, T. Damour and V. Mukhanov, Phys. Lett. B {\bf 458}, 209 (1999) ;
C. Armendariz-Picon, V. Mukhanov and P.J. Steinhardt, Phys. Rev. D{\bf 63}, 103510 (2001);
T. Chiba, T. Okabe and M. Yamaguchi, Phys. Rev. D{\bf 62}, 023511 (2000).


\bibitem{tachyonic} T. Padmanabhan, Phys. Rev. D{\bf 66}, 021301(R) (2002);
J.S. Bagla, H.K. Jassal, and T. Padmanabhan, Phys. Rev. D{\bf 67}, 063504 (2003).

\bibitem{holographic} M. Li, Phys. Lett. B {\bf 603} 1 (2004);
Q.G. Huang and M. Li, JCAP {\bf 0408}, 013 (2004);
Q.G. Huang and M. Li, JCAP {\bf 0503}, 001 (2005);
Q.G. Huang and Y.G. Gong, JCAP {\bf 0408}, 006 (2004);
X. Zhang and F.Q. Wu, Phys. Rev. D {\bf 72}, 043524 (2005);
M. Li, X.D. Li, S. Wang and X. Zhang, JCAP {\bf 0906} 036 (2009);
M. Li, X.D. Li, S. Wang, Y. Wang and X. Zhang, JCAP {\bf 0912} 014 (2009);
Y.T. Wang and L.X. Xu, Phys. Rev. D {\bf 81}, 083523 (2010).

\bibitem{agegraphic}
R.G. Cai, Phys. Lett. B {\bf 657}, 228 (2007);
H. Wei and R.G. Cai, Phys. Lett. B {\bf 660}, 113 (2008).


\bibitem{hessence} H. Wei, R.G. Cai, and D.F. Zeng, Class. Quant. Grav. {\bf 22}, 3189 (2005);
H. Wei, and R.G. Cai, Phys. Rev. D{\bf 72}, 123507 (2005).

\bibitem{Chaplygin}
A.Y. Kamenshchik, U. Moschella and V. Pasquier, Phys. Lett. B {\bf 511} 265 (2001);
M.C. Bento, O. Bertolami and A.A. Sen, Phys. Rev. D {\bf 66} 043507, (2002);
X. Zhang, F.Q. Wu and J. Zhang, JCAP {\bf 0601}, 003 (2006).


\bibitem{YMC} Y. Zhang, T.Y. Xia, and W. Zhao, Class. Quant. Grav. {\bf 24}, 3309 (2007);
T.Y. Xia and Y. Zhang, Phys. Lett. B {\bf 656}, 19 (2007);
S. Wang, Y. Zhang and T.Y. Xia, JCAP {\bf 10}, 037 (2008);
S. Wang and Y. Zhang, Phys. Lett. B {\bf 669}, 201(2008).

\bibitem{Union2} R. Amanullah et al., arXiv:1004.1711, ApJ accepted.

\bibitem{WeiHao} H. Wei, JCAP {\bf 1008}, 020 (2010);
S.F. Daniel et al., Phys. Rev. D. {\bf 81}, 123508 (2010);
L.X. Xu and Y.T. Wang, Phys. Rev. D {\bf 82}, 043503 (2010);
P.X. Wu and H.W. Yu, arXiv:1006.0674;
Y.G. Gong, X.M. Zhu, and Z.H. Zhu, arXiv:1008.5010.

\bibitem{YunWang} Y. Wang and K. Freese, Phys. Lett. B {\bf 632}, 201 (2006).


\bibitem{WMAP7} E. Komatsu et al., arXiv:1001.4538.


\bibitem{BAO} W.J. Percival et al., Mon. Not. Roy. Astron. Soc. {\bf 401} 2148 (2010).


\bibitem{Binned} D. Huterer and G. Starkman, Phys. Rev. Lett. {\bf 90}, 031301 (2003);
D. Huterer and A. Cooray, Phys. Rev. D {\bf 71}, 023506 (2005);
S. Sullivan, A. Cooray and D. E. Holz, JCAP {\bf 0709}, 004 (2007);
S. Qi, F. Y. Wang and T. Lu, Astron. Astrophys {\bf 483}, 49 (2008);
S. Qi, F. Y. Wang and T. Lu, Astron. Astrophys {\bf 487}, 853 (2008);
M. Kowalski et al., ApJ. {\bf 686}, 749 (2008).

\bibitem{Binned2} Y. Wang, arXiv:0904.2218, MPLA accepted.

\bibitem{Huang} Q.G. Huang, M. Li, X.D. Li and S. Wang, Phys. Rev. D {\bf 80}, 083515 (2009).

\bibitem{YunWangParametrization} Y. Wang and M. Tegmark, Phys. Rev. Lett. {\bf 92}, 241302 (2004);
Y. Wang and P. Mukherjee, Phys. Rev. D {\bf 76}, 103533 (2007); Y.
Wang, Phys. Rev. D {\bf 80}, 123525 (2009).

\bibitem{Simon} J. Simon, L. Verde and R. Jimenez, Phys. Rev. D {\bf 71}, 123001
(2005); T. Koivisto and D.F. Mota. Phys. Rev. D {\bf 73}, 083502
(2006); E.F. Martinez and L. Verde, JCAP {\bf 08}, 023 (2008).


\bibitem{swang} M.X. Lan, M. Li, X.D. Li and S. Wang, Phys. Rev. D {\bf 82}, 023516, (2010);
M. Li, X.D. Li and S. Wang, arXiv:0910.0717;
S. Wang, X.D. Li and M. Li, Phys. Rev. D {\bf 82}, 103006 (2010).

\bibitem{Liddle}
A.R. Liddle, MNRAS. {\bf 351}, L49 (2004);
M. Biesiada, JCAP. {\bf 02}, 003 (2007).

\bibitem{Schwarz}
G. Schwarz, Annals of Statistics. {\bf 6}, 461 (1978).

\bibitem{Perivolaropoulos} L. Perivolaropoulos, Phys. Rev. D {\bf 71}, 063503 (2005);
S. Nesseris and L. Perivolaropoulos, Phys. Rev. D {\bf 72}, 123519 (2005);
S. Nesseris and L. Perivolaropoulos, JCAP.  {\bf 0702}, 025 (2007).

\bibitem{Union Web} http://supernova.lbl.gov/Union/

\bibitem{Hu:1995en}
W. Hu and N. Sugiyama, ApJ {\bf 471}, 542 (1996).

\bibitem{Bond97}
J.R. Bond, G. Efstathiou and M. Tegmark, Mon. Not. R. Astron. Soc
{\bf 291}, L33 (1997).

\bibitem{BAODefzd} D.J. Eisenstein and W. Hu, ApJ. {\bf 496}, 605 (1998).

\bibitem{Eisenstein} D.J. Eisenstein et al., ApJ {\bf 633}, 560 (2005).

\bibitem{Hicken2} M. Hicken et al., ApJ {\bf 700}, 1097 (2009).

\end{thebibliography}
\end{document}